\documentclass[12pt]{article}

\usepackage{amsmath,amsfonts,epsfig,color,latexsym}

\newcommand{\be}{\begin{equation}}
\newcommand{\ee}{\end{equation}}
\newcommand{\bea}{\begin{eqnarray}}
\newcommand{\eea}{\end{eqnarray}}

\begin{document}

\begin{flushright}
hep-th/0410137\\
BROWN-HET-1425
\end{flushright}
\vskip.5in

\begin{center}

{\LARGE\bf On fuzzy spheres and (M)atrix actions }
\vskip 1in
\centerline{\Large Horatiu Nastase}
\vskip .5in

\end{center}
\centerline{\large  Brown University}
\centerline{\large Providence, RI, 02912, USA}

\vskip 1in

\begin{abstract}

{\large In this note we compare even and odd fuzzy sphere  constructions,
their dimensional reductions and possible (M)atrix actions having them as 
solutions. We speculate on how the fuzzy 5-sphere might appear as a solution 
to the pp wave (M)atrix model.
}

\end{abstract}

\newpage

Fuzzy spheres are of physical relevance (beyond their interesting mathematical
aspects) because of the possibility that they appear as solutions to (M)atrix
theory. As such, they give a possible quantum version of classical sphere 
solution, one which might be of relevance in the early Universe physics 
(see e.g. \cite{dkv}), as well as providing possible vacuum solutions in 
certain backgrounds. The fuzzy $S^2$ in fact is a solution to the M theory 
Matrix model in the maximally supersymmetric pp wave background \cite{bmn}.

We will therefore review some of the aspects of fuzzy spheres constructions, 
simplifying the analysis of odd spheres. We will see that in the described 
context, dimensional reduction of spheres becomes easier to understand (even
if still not straightforward). We will write down Matrix model actions
that have the even and odd spheres as solutions and argue that such an 
action will probably describe the quantum corrected version of the pp wave 
(M)atrix model, thus giving the conjectured 5-brane solution \cite{bmn}. 

\vspace{1cm}

{\em Fuzzy $S^2$}

The best understood case of fuzzy sphere is the fuzzy $S^2$. One needs 
noncommutative coordinates (realized by infinite matrices in the (M)atrix 
theory case) that satisfy an SO(3)-invariant algebra, generalizing the 
classical sphere. The algebra is  

\be
[X^i, X^j]=iR \epsilon^{ijk} X^k;\;\;\; (X^i)^2=R^2
\ee

By multiplying with $\epsilon^{ijk}$, the defining algebra becomes 
\be
\epsilon^{ijk} X^i X^j= iR X^k
\ee
which is equivalent to the previous.

\vspace{1cm}

{\em Fuzzy $S^4$} 

The case of fuzzy $S^4$ was analyzed in \cite{clt} following earlier work in 
\cite{gkp,kt}. Given the knowledge 
that it had to be a solution to the (M)atrix action carrying 
4-brane charges, the authors defined the algebra to be satisfied as 
(again, a suitable SO(5)-invariant extension of the classical 4-sphere)
\bea
&&\epsilon^{ijklm} X_iX_j X_k X_l=\alpha X^m\nonumber\\
&& (X^i)^2=R^2\nonumber\\
&& R_{ij} X_j=U(R) X_i U(R^{-1})
\label{foursph}
\eea
It is not clear that this definition is equivalent to
\be
[X_i, X_j]=\beta \epsilon_{ijklm}X^kX^lX^m
\label{foursphtoo}
\ee
which would be another possible definition of the fuzzy $S^4$. As we noted, 
in the $S^2$ case the two definitions are identical.

The explicit construction of the fuzzy $S^4$ though does not cover all possible
4-brane charges, only those that can be written as
\be
N=\frac{(n+1)(n+2)(n+3)}{6}
\ee
In that case, the construction is 
\be
X^i=\sum_r\rho_r(\Gamma^i);\;\;\; X^i:(V^{\otimes n})_{sym}\rightarrow
(V^{\otimes n})_{sym}
\ee
where $\rho_r(X)$ inserts X on the r position in $V^{\otimes n}$, and V is the
vector space of spinor representation for SO(5). This explicit construction 
satisfies both (\ref{foursph}) and (\ref{foursphtoo}).

The explicit construction of the 
fuzzy $S^4$ also satisfies the equations of motion
\be
[X^j, [X^i, X^j]]+aX^i=0
\label{eom}
\ee
since $J^{ij}=[X^i,X^j]=\sum_r\rho_r([\Gamma_i, \Gamma_j])$ 
are the generators of SO(5). But it is not clear if these equations of 
motion are a 
consequence of one of the original forms of the algebra (\ref{foursph}) or
(\ref{foursphtoo}). The equations of motion in (\ref{eom})  
come from the action
\be
S=\int \frac{[X^i,X^j]^2}{2}+a(X^i)^2
\ee

For $S^2$, these equations of motion can also be rewritten as 
\be
[J_{ij}, X_j]=4X_i
\ee
since $J_{ij}\propto [X_i, X_j]$, except that now one can 
easily see that the fuzzy $S^2$ algebra implies these equations of motion. 

\vspace{1cm} 
{\em Fuzzy $S^{2k}$}

The construction generalizes easily to even spheres $S^{2k}$ as 
(the detailed analysis was done in \cite{ram,hr} and further clarified 
in \cite{kimura,ab})
\bea
&&\epsilon^{i_1...i_{2k}} X_{i_1}... X_{i_{2k-1}}=\alpha X^{i_{2k}}\nonumber\\
&& (X^i)^2=R^2\nonumber\\
&& R_{ij} X_j=U(R) X_i U(R^{-1})
\eea
or (equivalently?) 
\be
[X_i, X_j]=\beta\epsilon_{iji_2...i_{2k}}X^{i_3}...X^{i_{2k}}
\ee
solved by the explicit construction $X_i=\sum_r\rho_r(\Gamma_i)$ 
that satisfies the equations of motion 
\be
[J_{ij}, X_j]=4k X_i
\ee
due to $J_{ij}\propto [X_i,X_j]$.

\vspace{1cm}
{\em Odd dimensional spheres: construction and algebra}

We are again looking for a matrix representation that generalizes the
coordinates $X^i$, and since we want a generalization of a sphere, we want 
to have $(X^i)^2=$ const. and an algebra (coming hopefully from some 
simple equations of motion) which are SO(2n)-invariant. The fuzzy 3-sphere 
was introduced in \cite{gr} and further developped and generalized to odd 
spheres in \cite{sanjaye}.

We will follow the analysis in \cite{sanjaye}, and we will focus on the 
$S^3$ case, leaving the generalization to any odd spheres to the end.

Take the vector space V of spinor representations 
for SO(4). It splits into positive and negative chirality representations, 
as $SO(4)=SU(2)\times SU(2)$, thus $V=V_++V_-$. Take the subspace 
${\cal R}_n$ of $Sym(V^{\otimes n})$. It will substitute $Sym(V^{\otimes n})$
as the basis vector space for the fuzzy sphere representation. ${\cal R}_n$ is 
defined as ${\cal R}_n^++{\cal R}_n^-$, where ${\cal R}_n^+$ and ${\cal R}_n^-$
are spaces of $SU(2)\times SU(2)$ labels $(\frac{n+1}{2}, \frac{n-1}{2})$ 
and $(\frac{n-1}{2}, \frac{n+1}{2})$ respectively (denoting the number of 
$V_+$ and $V_-$ factors in each).

Then 
\be
X^i= {\cal P}_{{\cal R}_n}\hat{X}_i{\cal P}_{{\cal R}_n}=
{\cal P}_{{\cal R}_n^+}\hat{X}_i{\cal P}_{{\cal R}_n^-}+
{\cal P}_{{\cal R}_n^-}\hat{X}_i{\cal P}_{{\cal R}_n^+}=X_i^++X_i^-
\ee
where ${\cal P}_{{\cal R}_n}={\cal P}_{{\cal R}_n^+}+{\cal P}_{{\cal R}_n^-}$ 
is the projector and 
\be
\hat{X}_i=\sum_r \rho_r(\Gamma_i)
\ee
Then $X_i=X_i^++X_i^-$ and $Y^i =X_i^+-X_i^-$ are two independent variables 
that need to be used to define the fuzzy $S^3$, or equivalently 
we can use $X_i^+, X_i^-$.

Following  \cite{sanjaye}, we can compute that $X_i^2$ commutes with the SO(4) 
generators and is indeed constant in this subspace, 
\be
X_i^2{\cal P}_{{\cal R}_n}=\frac{(n+1)(n+3)}{2}\equiv N
\ee
For a fuzzy $S^{2k-1}$, one gets $(n+1)(n+2k-1)/2$ on the r.h.s.

One needs to take the irrep ${\cal R}_n$  as opposed to $Sym(V^{\otimes n})$
in the even sphere case, and then 
$X_i^2$ is constant 
on the space. In the calculation, one needs to be careful, since
\be
\sum_i{\cal P}_{{\cal R}_n} \sum_r \rho_r(\Gamma_i)\sum_s \rho_r(\Gamma_i)
{\cal P}_{{\cal R}_n}\neq \sum_i X_i^2
\ee

For the definition of the fuzzy 3-sphere algebra, \cite{sanjaye} 
defines the objects 
\bea
X_i^+= {\cal P}_{{\cal R}_n^-}\sum_r \rho_r(\Gamma_i P_+)
{\cal P}_{{\cal R}_n^+}\;\;\;\;&&
X_i^-= {\cal P}_{{\cal R}_n^+}\sum_r \rho_r(\Gamma_i P_-)
{\cal P}_{{\cal R}_n^-}\nonumber\\
X_{ij}^+= {\cal P}_{{\cal R}_n^+}\sum_r \rho_r(\frac{1}{2}[\Gamma_i,
\Gamma_j] P_+)
{\cal P}_{{\cal R}_n^+}\;\;\;\;&&
Y_{ij}^+= {\cal P}_{{\cal R}_n^+}\sum_r \rho_r(\frac{1}{2}[\Gamma_i, 
\Gamma_j] P_-)
{\cal P}_{{\cal R}_n^+}\nonumber\\
X_{ij}^-= {\cal P}_{{\cal R}_n^-}\sum_r \rho_r(\frac{1}{2}[\Gamma_i,
\Gamma_j] P_-)
{\cal P}_{{\cal R}_n^-}\;\;\;\;&&
Y_{ij}^-= {\cal P}_{{\cal R}_n^-}\sum_r \rho_r(\frac{1}{2}[\Gamma_i,
\Gamma_j] P_+)
{\cal P}_{{\cal R}_n^-}\nonumber\\
\label{obj}
\eea
where $P_{\pm}$ are projectors onto $V_{\pm}$ and then also 
\bea
X_i=X_i^++X_i^-\;\;\;\; &&
Y_i=X_i^+-X_i^-\nonumber\\
X_{ij}=X_{ij}^++X_{ij}^-\;\;\;\;&&
Y_{ij}=Y_{ij}^++Y_{ij}^-\nonumber\\
\tilde{X}_{ij}=X_{ij}^+-X_{ij}^-\;\;\;\;&&
\tilde{Y}_{ij}=Y_{ij}^+-Y_{ij}^-
\label{def}
\eea
where all $X_{ij}^+$ is selfdual ($X_{ij}^+=1/2 \epsilon_{ijkl}X_{kl}^+$)
and so is $Y_{ij}^-$, whereas $X_{ij}^-$ and $Y_{ij}^+$ are anti-selfdual.
The algebra is then defined by a large series of (anti)commutators between 
these generators.

However, one can easily check that that algebra implies
\bea
&& (n+3) X_{ij}= [X_i,X_j]+ \frac{1}{2}\epsilon_{ijkl}\{ X_k, Y_l\};\;\;
X_{ij}=\frac{1}{2}\epsilon_{ijkl}\tilde{X}_{kl}
\nonumber\\
&&-(n+1)Y_{ij}=[X_i,X_j]- \frac{1}{2}\epsilon_{ijkl}\{ X_k, Y_l\};\;\;
Y_{ij}=-\frac{1}{2}\epsilon_{ijkl}\tilde{Y}_{kl}
\nonumber\\
&& (n+3) \tilde{X}_{ij}=\{ X_i, Y_j\} + \epsilon_{ijkl} X_k X_l\nonumber\\
&&-(n+1)\tilde{Y}_{ij}=\{ X_i, Y_j\} - \epsilon_{ijkl} X_k X_l
\eea
and thus only $X_i$ and $Y_i$ parametrize the fuzzy algebra, subject to the 
constraint
\be
[X_i, X_j]=-[Y_i, Y_j]
\ee
Since ${\cal P}_{{\cal R}_n^-}{\cal P}_{{\cal R}_n^+}={\cal P}_{{\cal R}_n^+}
{\cal P}_{{\cal R}_n^-}=0$, $X_i^+X_j^-=X_i^-X_j^+=0$, and thus 
$X_i^2=X_i^+X_i^-+X_i^-X_i^+=-Y_i^2=N$.

From the explicit form in (\ref{def}) one can find that the SO(4) generators
are 
\be
J_{ij}\equiv {\cal P}_{{\cal R}_n}\sum_r \rho_r (\frac{1}{2}[\Gamma_i, 
\Gamma_j]){\cal P}_{{\cal R}_n}= X_{ij}+Y_{ij}
\ee
and so 
\be
J_{ij}= -\frac{2}{(n+1)(n+3)}[X_i, X_j]+\frac{(n+2)}{(n+1)(n+3)}
\epsilon_{ijkl}\{ X_k , Y_l\}
\equiv A[X_i, X_j]+B\epsilon_{ijkl}\{ X_k , Y_l\}
\ee
where $A=-1/N$ and $B=(n+2)/(2N)= \sqrt{2N+1}/(2N)$.

With this definition, the only independent equation of motions, giving the 
algebra of the fuzzy sphere in terms of $X_i$, $Y_i$, are 
the rotation properties
of $X_i, Y_i$, namely
\be
[J_{ij}, X_j]=6X_i;\;\;\;[J_{ij}, Y_j]=6Y_i
\ee
where of course one has to replace $J_{ij}$ with its explicit form as a 
function of $X_i, Y_i$, and always subject to the contraint $[X_i, X_j]
=-[Y_i, Y_j]$. This is quite satisfying, given that the even fuzzy sphere 
equations were also related to rotational invariance (there, through 
the presence of the epsilon tensor). 

One has to find actions that have these equations of motion. But first notice 
that $Y_i$ is anti-hermitian, thus we need to redefine $Y_i=i\tilde{Y}_i$
and thus find an action for $X_i, \tilde{Y}_i$ modulo the constraint
$[X_i,X_j]=[\tilde{Y}_i, \tilde{Y}_j]$. 

\vspace{1cm}
{\em Actions}

One can easily check that such an action is (one writes such an action with 
arbitrary coeffficients for all the terms and then fixes the coefficients by 
requiring to get the correct equations of motion)
\bea
&&L=\frac{1}{2N}Tr\{ \frac{1}{4}([X_i,X_j]^2-[\tilde{Y}_i, \tilde{Y}_j]^2) 
-3N X_i^2 +3N \tilde{Y}_i^2+\nonumber\\
&&i \frac{\sqrt{2N+1}}{2}\epsilon_{ijkl} (X_i X_j \{ X_k, \tilde{Y}
_l\}+\tilde{Y}_i \tilde{Y}_j\{ \tilde{Y}_k, X_l\})\}
\eea
One notes that the $\tilde{Y}$ terms have the wrong sign.
Thus the energy of a static solution (so that $E=-L$)
 satisfying $[X_i, X_j]=[\tilde{Y}_i,
\tilde{Y}_j]$ and $X_i^2=\tilde{Y}_i^2$ (constraints) is easily seen to be 
zero.
The same was true  for the fuzzy $S^2$ solution to the pp wave (M)atrix model
in \cite{bmn},
except that in that case we needed
the actual equations of motion, not just a constraint.

The construction of the algebra and action for the fuzzy $S^5$ case is 
similar. One can still define the objects in (\ref{obj},\ref{def}) 
subject to the 
constraint that $[X_i,X_j]=-[Y_i,Y_j]$. We will get similar equations of 
motion (by simplifying the algebra). Without doing the explicit calculations,
from SO(6) invariance and using the constraint, we can say that  
the SO(6) generator is
\be
J_{ij}=X_{ij}+Y_{ij}=-a[X_i, X_j]+b\epsilon_{ijklmn} X_k X_l \{X_m , Y_n\}
\ee
(where a and b could be determined by explicit calculation), 
with equations of motion 
\be
[X_{ij}+Y_{ij}, X_i]=10 X_i;\;\;\; [X_{ij}+Y_{ij}, Y_j]=10 Y_j
\ee

Then the corresponding action is (after redefining $Y_i=i\tilde{Y}_i$)
\bea
&&L=Tr\{ \frac{a}{4}([X_i,X_j]^2-[\tilde{Y}_i,\tilde{Y}_j]^2)
-5(X_i^2-\tilde{Y}_i^2)\nonumber\\
&&+i\frac{b}{2} \epsilon_{ijklmn}(X_iX_jX_kX_l\{X_m, \tilde{Y}_n\}+
\tilde{Y}_i\tilde{Y}_jX_kX_l \{\tilde{Y}_m, X_n\}) \}
\label{fuzfive}
\eea

This construction  generalizes also easily to $S^{2k-1}$. 
Using similar arguments as for the $S^5$ case, the 
generator of $SO(2k)$ rotations is 
\be
J_{ij}=X_{ij}+Y_{ij}=-a[X_i, X_j]+b\epsilon_{iji_3...i_{2k}}X_{i_3}...
X_{i_{2k-2}}\{X_{i_{2k-1}}, Y_{i_{2k}}\}
\ee
and the equations of motion are 
\be
[X_{ij}+Y_{ij}, X_j]=2(2k-1)X_i;\;\;\;[X_{ij}+Y_{ij}, Y_j]= 2(2k-1)Y_j
\ee
They come from the action 
\bea
&& L=Tr\{ \frac{a}{4}([X_i,X_j]^2-[\tilde{Y}_i, \tilde{Y}_j]^2)-(2k-1)
(X_i^2-\tilde{Y}_i^2)\nonumber\\
&&+i\frac{b}{2}
 \epsilon_{i_1...i_{2k}}(X_{i_1}...X_{i_{2k-2}}\{X_{i_{2k-1}}, Y_{i_{2k}}
\}+Y_{i_1}Y_{i_2}X_{i_3}...X_{i_{2k-2}}\{Y_{i_{2k-1}}, X_{i_{2k}}\})
\eea

A simple observation is that the action
\be
([X^i,X^j]-[Y^i,Y^j])^2
\ee
also has (all) 
the odd fuzzy spheres as a trivial solution, since both equations 
of motion are proportional to the constraint $[X^i,X^j]-[Y^i,Y^j]=0$. 

More realistically, adding 
\be
\alpha ([X^i,X^j]-[Y^i,Y^j])^2
\ee
to the previous action still satisfies the odd 
fuzzy sphere equations of motion, so there is a one-parameter set of actions 
with the fuzzy odd spheres as solutions. 

\vspace{1cm}
{\em Dimensional reduction}

For classical spheres, the ``diameter'' of a sphere (reached when one of 
the coordinates takes an extreme value- maximum or 0) is a sphere of 
one less dimension. In fact, for any fixed value of one of the coordinates,
we get a lower dimensional sphere.
Let's check whether we can obtain the same for fuzzy spheres. 

The simplest case is the case of applying the procedure twice, for 
embedding even spheres into even spheres. For n=2k+1, embedding $S_{n-1}$ into 
$S_{n+1}$ can be done at the level of the algebra:
\bea
&&\sum _{i=1}^n \hat{X}_i^2 =1-\hat{X}_{n+1}^2-\hat{X}_{n+2}^2\nonumber\\
&&[\hat{X}_i,\hat{X}_j]= \epsilon_{iji_3....i_ni_{n+1}i_{n+2}}\hat{X}^{i_3}...
\hat{X}^{i_n}(\hat{X}^{i_{n+1}}\hat{X}^{i_{n+2}})
\eea
and thus if the $S_{n+1}$ operators $[\hat{X}^{i_{n+1}}, \hat{X}^{i_{n+2}}]$,
$ \hat{X}_{i_{n+1}}^2$, $\hat{X}_{i_{n+2}}^2$
have given eigenvalues, the $S_{n+1}$ algebra reduces to the $S_{n-1}$ algebra.

Let us now try to embed $S^3$ into $S^4$, not the explicit representation, 
but using only the algebra. As we saw, for the even dimensional spheres, 
we can represent the 5d operators $\hat{X}_{\mu}$ by $\Gamma_{\mu}$.
In particular, for dimensional reduction, we can represent $\hat{X}_5=
\Gamma_5$ (more precisely, $\hat{X}_5=\sum_r \rho_r(\Gamma_5)$).

Since we have 
\be
[\Gamma_i, \Gamma_j]=\epsilon_{ijkl5}\Gamma_k\Gamma_l \Gamma_5=
\frac{1}{2}\epsilon_{ijkl}\{ \Gamma_k, \Gamma_l \Gamma_5\}
\ee
we see that a good way to set up the dimensional reduction of the 
fuzzy 4-sphere algebra is 
\be
[\Hat{X}_i, \hat{X}_j]=\frac{1}{2}
\epsilon_{ijkl}\{ \hat{X}_k, \hat{X}_l\hat{X}_5\}
\label{algebra}
\ee

Next, notice that if we dimensionally reduced as follows: $\hat{X}_i=X_i, 
\hat{X}_i \hat{X}_5= Y_i$ as one would think natural, then we would get 
$Y_{ij}=0, X_{ij}=2[X_i,X_j]$, which is not what we want (we need to have
two independent sets of variables). 

Fortunately, we can see from the explicit gamma matrix representation that 
this is not quite so. In fact, 
\be
X_i=X_i^++X_i^-= {\cal P}_-\sum_r\rho_r (\Gamma_i P_+){\cal P}_+
+ {\cal P}_+\sum_r\rho_r (\Gamma_i P_-){\cal P}_-
\ee
where $P_{\pm}= (1\pm \Gamma_5)/2$ are the projectors onto given chiralities.
That still means that $X_i^{\pm}\hat{X}_5=\pm X_i^{\pm}$, so that 
$X_i\hat{X}_5=Y_i=-\hat{X}_5 X_i$.  
For the dimensional reduction ansatz, we would need to write something like
\be
X_i= {\cal P}_-\hat{X}_i\frac{1+\hat{X}_5}{2}{\cal P}_++
{\cal P}_+\hat{X}_i\frac{1-\hat{X}_5}{2}{\cal P}_-
\ee
and now (\ref{algebra}) will not be true without hats anymore. Rather, without 
hats, the two sides of (\ref{algebra}) will be independent.  
This form of the dimensional reduction is not very appealing, 
as we still need the projectors
${\cal P}_{\pm}$ which refer to the explicit representation of the X's as 
gamma matrices, but we can find no better way of doing it.

In any case, then one has to replace this definition in 
\be
[\hat{J}_{ij}, \hat{X}_j]=6 \hat{X}_i
\ee
(the 6 instead of 8 is because the summation is restricted: we don't sum 
over the 5th coordinate) 
and using exactly the same calculation as was done with the gamma matrices 
(except that now we don't use that notation), obtain that it dimensionally 
reduces to the desired 
\be
[J_{ij}, X_j]= 6X_i;\;\;\; J_{ij}=A[X_i,X_j]+B\epsilon_{ijkl}
\{X_k, Y_l\}
\ee
where of course $Y_i= X_i\hat{X}_5$. 

For the reverse dimensional reduction, of odd sphere to even sphere, 
e.g. $S_3$ to $S_2$, we again start from the dimensional reduction of the 
representation, to gain insight about the dimensional reduction of the 
algebra. We have 
\be
\hat{X}_{\mu}={\cal P}_{{\cal R}_n}\sum_r \rho_r (\hat{\Gamma}_{\mu})
{\cal P}_{{\cal R}_n},\;\;\; \mu=1,4
\ee
with the dimensional reduction of the gamma matrices (easily generalizable 
to any odd sphere)
\be
\hat{\Gamma}_i=\Gamma_i\otimes \sigma_1;\;\;
\hat{\Gamma}_4=1\otimes \sigma_2;\;\;
\hat{\Gamma}_5= 1\otimes \sigma_3
\ee
Then put 
\be
\hat{Y}_4=\sum_r\rho_r(\hat{\Gamma}_4\hat{\Gamma}_5)=\sum_r\rho_r (1\otimes
\sigma_1)\Rightarrow \hat{X}_i\hat{Y}_4=\sum_r\rho_r (\Gamma_i)\equiv 
X_i
\ee
With this dimensional reduction ansatz, we can write down the dimensional 
reduction of the SO(4) rotation generator,
\be
\hat{J}_{ij}= a[\hat{X}_i,\hat{X}_j]+b\epsilon_{ijk4}\{\hat{X}_k, \hat{Y}_4\}
= a[\hat{X}_i,\hat{X}_j]+b\epsilon_{ijk}X_k
\ee
and thus deduce the dimensional reduction of the equations of motion as 
follows. If we sandwich $[\hat{J}_{ij}, \hat{X}_j]= 4 \hat{X}_i$ between two 
$\hat{Y}_4$'s  we get 
\be
[J_{ij}, X_j]= 4X_i;\;\;\; J_{ij}= a[X_i,X_j]+b\epsilon_{ijk}X_k
\ee
It seems though that one still needs to impose the more restrictive equation 
of motion $[X_i, X_j]= \epsilon_{ijk}X_k $ to get the correct dimensional 
reduction.

\vspace{1cm}
{\em PP wave Matrix model and fuzzy 5-sphere}

We expect the fuzzy 5-sphere to be a solution to of the pp wave Matrix 
Model in \cite{bmn}.

In the supergravity description of M theory, there are two types of 
giant gravitons  in the pp wave background. There are 
two-spheres that correspond to
giant gravitons in $AdS_4$ or $S_4$ before the Penrose limit, and
whose radius is 
\be
r= \frac{\pi}{6}\mu p^+= \frac{\pi}{6}\frac{\mu N}{R}
\ee
(the second line corresponds to the naive Matrix theory DLCQ) 
and there are also 5-spheres, that correspond to giant gravitons in $AdS_7$
or $S_7$ before the Penrose limit, whose radius obeys
\be
r^4= \frac{8\pi^2}{3} \mu p^+= \frac{8\pi^2}{3}\frac{\mu N}{R}
\ee

But in the Matrix model in \cite{bmn} 
there are exactly two classical vacuum solutions, a 
fuzzy two-sphere,
\be
[\phi^i,\phi^j]=i\frac{\mu}{6R}\epsilon_{ijk} \phi^k;\;\;\;
r=2\pi \sqrt{\frac{Tr[\sum_i {\phi^i}^2]}{N}}
\ee
and the vacuum, $\phi^i=0$. The fuzzy two-sphere solution of the Matrix model
has the correct (supergravity) radius, as expected.

The Matrix model is written with its coupling $1/g^2$ in front
($g=(R/(\mu N))^{3/2}$), and then 
we have for the two-sphere in the rescaled variables
$\hat{\phi}\sim \mu/g$ (so the solution is 
classical, $g\hat{\phi}\sim o(1)$,
as we have checked). But for the 5-sphere giant graviton, we would obtain 
$\hat{\phi}^4\sim \mu/g$, 
which therefore is quantum in nature in the Matrix model.
So the only candidate, the $\phi=0$ vacuum should quantum mechanically be 
blown up to a 5-sphere, which then should get the correct radius.

Indeed, \cite{juan} have shown that the linear fluctuation spectrum of a
5-brane matches (exactly) with the protected spectrum of excited states 
about the X=0 vacuum of the Matrix model (found by computing small QM 
fluctuations and symmetry arguments).

So we would want to obtain the fuzzy 5-sphere as a solution of the exact 
quantum-mechanically corrected Matrix model.

There are two ways in which it seems possible to do this using our description
of the fuzzy 5-sphere. One would be to 
embed both $X_i$ and $Y_i$ into the $\phi_i$ of the Matrix model. But we 
can easily check that this is not possible for the action in (\ref{fuzfive}).
We could add 
\be
\alpha ([X^i,X^j]-[Y^i,Y^j])^2
\ee
to the action, obtaining (for $\alpha =a$)
\bea
L&=&Tr\{ \frac{a}{2}([X_i,X_j]^2-[X_i,X_j] [\tilde{Y}_i,\tilde{Y}_j])
-5(X_i^2-\tilde{Y}_i^2)\nonumber\\
&&+i\frac{b}{2} \epsilon_{ijklmn}(X_iX_jX_kX_l\{X_m, \tilde{Y}_n\}+
\tilde{Y}_i\tilde{Y}_jX_kX_l \{\tilde{Y}_m, X_n\}) \}
\eea
But it is not clear how we would go about embedding this either.

We should note here that whenever we write a fuzzy sphere action as a Matrix 
action we implicitly assume that the coupling has been extracted as an overall
$1/g^2$, so that we can get back the coupling dependence by rescaling fields 
and coordinates.

The only other way (without adding the extra term to the action)
seems to be to have $Y^i$ being a bilinear in the fermion 
fields, of the type $\Psi\Gamma^i\Psi$.

This is not so unusual. For instance, the QCD chiral symmetry breaking phase 
transition order parameter is believed to be $<\bar{q}q>$ (the quark 
condensate). Also, in the case of the Seiberg \cite{seiberg} and Seiberg-Witten
\cite{sw} analysis of the ${\cal N}=1$ and ${\cal N}=2$ susy gauge theories,
the order parameter of the chiral symmetry phase breaking is the ``meson''
field $M=\tilde{Q}Q$, that acquires a nonzero VEV. In both cases, the 
nonzero VEV of the bilinear in fields appears nonperturbatively, and is 
essential to the physics. 

Therefore it is not unlikely to have a quantum theory with an ``effective 
potential'' for the ``meson'' field $\Psi \Gamma^i \Psi$, of the needed form.
It is unfortunately not clear how to calculate it. 
The only difference from the meson case is that now we want to have an 
object with gauge indices (not a gauge invariant combination), but given 
that fields are in the adjoint representation of the gauge group, it is 
the natural thing to do. 

There is a simple way to argue for the presence of the fuzzy 5-sphere action.
Myers \cite{myers} derived a term $F_{tijk}Tr(X^i
X^j X^k)$ that boosted gave  $F_{+ijk}Tr(X^iX^jX^k)=\epsilon_{ijk}Tr(X^iX^j
X^k)$ for the pp wave (M)atrix model. 
 It came from the classical DBI coupling $\int i_X (C_{(3)})$.

If we would have a dual $C_{(6)}$ field in the same background,
it should similarly provide a term
\be
F_{+ijklmn}Tr(X^iX^jX^kX^lX^mX^n)= \epsilon_{ijklmn}Tr(X^iX^jX^kX^lX^mX^n)
\ee

Notice then that a nonlinear susy transformation
\bea
&&\delta X_n= \Psi^T \Gamma_n \epsilon\nonumber\\
&&\delta \Psi= \epsilon
\eea
would relate it to a term of the desired form
\be
\epsilon_{ijklmn}Tr(X^iX^jX^kX^l\{ X^m, \Psi^T\Gamma^n \Psi\})
\ee

Then this term, together with the usual commutator and mass terms, 
\be
[X^i,X^j]^2-m^2 (X^i)^2
\ee
provide half of the terms in the fuzzy 5-sphere action, and the other half 
are there to cancel the energy when the constraints are satisfied. 
As the action is valid only on the constraints anyway, this seems 
enough to argue for the plausibility of the action.

So we have found that the action (\ref{fuzfive}) is a possible candidate 
for describing the quantum 5-sphere solution of the pp wave Matrix model.
This action is unique if we impose the equations of motion and the condition 
of minimal corrections to the Matrix model, so this gives it a certain 
degree of plausibility.

We might be worried that the action would have negative energy configurations,
but we have to remember that the action is only valid if the constraints are 
satisfied, and then the energy is automatically zero. 

We  observe that the action (\ref{fuzfive}) has also $X=Y=0$ as a solution
(along with many others),
and thus why would we call the fuzzy 5-sphere solution a blow-up of the 
X=0 solution of the BMN Matrix model? We only gave an intuitive argument 
for half of the terms in the action, the rest were guessed by using 
supersymmetry and the constraints. In reality, there should probably 
be more terms for $Y_i=\Psi^T \Gamma^i\Psi$ which would force it to be nonzero,
and for a solution satisfying our constraints that would mean the X's 
would be nonzero too. 

Note that \cite{sheikh} has also proposed an action that would have the fuzzy 
3-sphere as a solution, different than the one proposed in this paper,
but that construction is slightly different in scope,
being motivated by a 3-brane discretization, and the fuzzy 3-sphere 
construction used there is different, using a constant matrix depending on 
the representation. 

Finally, we should note that the fuzzy $S^4$ construction does not cover 
all possible 4-brane charges, and one needs something else (maybe 
nonassociativity) to describe the general case. Similarly, it is possible 
that the fuzzy 5-sphere construction will only provide certain cases for the 
expected giant 5-sphere graviton.

{\bf Acknowledgements} I thank Sanjaye Ramgoolam for many discussions and 
for collaboration at the early stages of this project.
I would also like to acknowledge useful discussions 
with Antal Jevicki and Jeff Murugan.
This research was  supported in part by DOE
grant DE-FE0291ER40688-Task A.

\end{document}